\documentclass[prb,aps,twocolumn,superscriptaddress,10pt]{revtex4-1}
\usepackage{amssymb,amsfonts,amsmath,graphicx,color,times,mathtools,dsfont}
\pdfoutput=1
\usepackage[normalem]{ulem}
\usepackage{hyperref}

\def\tr{\mbox{tr}}
\def\bra#1{\langle{#1}|}
\def\ket#1{|{#1}\rangle}

{\catcode`\|=\active 
  \gdef\Braket#1{\begingroup
\mathcode`\|32768\let|\BraVert\left<{#1}\right>\endgroup}}
\def\BraVert{\egroup\,\mid\,\bgroup}

\DeclareMathOperator{\rank}{rank}

\newcommand{\T}{T}

\newcommand{\e}{\mathrm{e}}
\newcommand{\1}{\mathds{1}}

\definecolor{christian}{rgb}{0.3,0.3,0.8}

\definecolor{francesca}{rgb}{0,0.5,0.9}

\definecolor{john}{rgb}{1,0.1,0.1}

\allowdisplaybreaks

\begin{document}

\title{Total correlations of the diagonal ensemble as a generic indicator for ergodicity breaking in quantum systems}

\author{F.\ Pietracaprina}
\email{Francesca.Pietracaprina@roma1.infn.it}
\affiliation{Dipartimento di Fisica, Universit\`a degli Studi di Roma ``La Sapienza'', I-00185, Roma, Italy}

\author{C.\ Gogolin}
\email{publications@cgogolin.de}
\affiliation{ICFO-The Institute of Photonic Sciences, Mediterranean Technology Park, 08860 Castelldefels (Barcelona), Spain}

\author{J.\ Goold}
\email{jgoold@ictp.it}
\affiliation{The Abdus Salam International Centre for Theoretical Physics (ICTP), Trieste, Italy}

\begin{abstract}
The diagonal ensemble is the infinite time average of a quantum state following unitary dynamics in systems without degeneracies. In analogy to the time average of a classical phase space dynamics, it is intimately related to the ergodic properties of the quantum system giving information on the spreading of the initial state in the eigenstates of the Hamiltonian. In this work we apply a concept from quantum information, known as {\it total correlations}, to the diagonal ensemble. Forming an upper-bound on the multipartite entanglement, it quantifies the combination of both classical and quantum correlations in a mixed state.
We generalize the total correlations of the diagonal ensemble to more general $\alpha$-Renyi entropies and focus on the the cases $\alpha=1$ and $\alpha=2$ with further numerical extensions in mind.  
Here we show that the total correlations of the diagonal ensemble is a generic indicator of ergodicity breaking, displaying a sub-extensive behaviour when the system is ergodic.
We demonstrate this by investigating its scaling in a range of spin chain models focusing not only on the cases of integrability breaking but also emphasize its role in understanding the transition from an ergodic to a many-body localized phase in systems with disorder or quasi-periodicity.   
\end{abstract}

\maketitle

\makeatletter

\section{Introduction}
Attempts to recover statistical mechanics from the underlying unitary dynamics of a quantum system are around since the inception of quantum theory with the pioneering approaches of both von Neumann~\cite{Vonneumann:1929} and Schr\"{o}dinger~\cite{Schrodinger:1927}.
Although these works showed impressive foresight, until relatively recently these foundational studies were almost forgotten and seen as irrelevant due to the fact that unitary evolution was not relevant over dynamical time scales in the laboratory.
Arguments for the validity of statistical mechanics predominantly consisted of invoking coupling to the larger bath of the universe and hence thermalisation by dissipation.

In the past two decades, these foundational questions have seen an unprecedented resurgence in interest by theorists from several different scientific communities, ranging from condensed matter physics to quantum information \cite{Zelevinsky1996,Polkovnikov:2011,Yukalov:2011,Eisert:2015,Goold:2016,Gogolin:2016,Borgonovi:2016,Alessio:2016}.
This revival is due, in no small part, to great advances in experimental ultra-cold atomic physics \cite{Lewenstein:2007} where pioneering experiments were successful in generating and probing coherent unitary dynamics over long timescales \cite{Greiner:2002,Kinoshita:2006,Trotzky:2012,Kaufman2016}.
This includes an experimental realization \cite{Kinoshita:2006} of the Lieb-Liniger model of interacting bosons in one dimension, in which the existence of an extensive set of conserved quantities, due to the integrability of the model, renderes the dynamics non-ergodic \cite{Rigol:2008}.

Experimental motivations aside, there has also been developments in theoretical condensed matter physics which have forced us to carefully think about the foundations of statistical mechanics beyond the paradigm of integrable systems.
In particular Basko, Aleiner and Altshuler have demonstrated that Anderson localisation \cite{Anderson:1958} is stable in the presence of interactions \cite{Basko:2006} leading to a new type of transition to a phase which is known as many-body localization (MBL) \cite{Nandkishore:2015,Vosk:2015,Altman:2015}.
Interestingly this transition is between an ergodic and a non-ergodic phase \cite{Oganesyan:2007} and has led to an intense interest in the phenomenology of ergodicity and its breaking in quantum dynamics.

Physically motivated, an approach known as Eigenstate Thermalisation Hypothesis (ETH) has proven to be popular amongst researchers from this community.
The ETH was born out of a realization by Berry \cite{Berry:1977}, who postulated that, in the semi-classical limit of quantum systems with chaotic classical counterparts, the Wigner function evaluated on eigenstates reduces to the micro-canonical distribution.
This was extended to arbitrary systems by Deutsch \cite{Deutsch:1991}, who proposed to assume that generic eigenstates of ergodic systems are like eigenstates of full random matrices.
Building on these ideas Srednicki formulated what is now known as the ETH, which is an ansatz on the behavior of matrix elements of observables with the consequence that ergodic systems can show thermal behavior at the level of individual eigenstates \cite{Srednicki:1994,Srednicki:1996,Srednicki:1999}.

Another approach, popular in the quantum information and mathematical physics communities, is the concept of normal and canonical typicality \cite{Tasaki:1998,Gemmer:2003, Popescu:2006,Goldstein:2006,Reimann:2008}.
This approach has the advantage of not just replacing one hypothesis (that systems tend to equilibrate to Gibbs states) with another equally unproven one (that systems generically fulfill the ETH), but it replaces the equal a priori probability postulate (all states in a microcanonical shell are equally probable) with a strictly weaker assumption, by showing rigorously, that the overwhelming majority of states in a microcanonical shell have nearly the same properties with respect to certain observables, like, for example, local ones.
However, the generality of the results obtained based on these concepts makes it difficult to apply this approach to concrete systems, as in realistic situations interesting dynamics is usually starting from a highly untypical initial condition.

In addition to the above, in recent years dynamical equilibration of expectation values and density matrices of subsystems under unitary dynamics has been studied extensively \cite{Reimann:2008,Linden:2009,Reimann10,Reimann12,Malabarba2014,Reimann2016} (see also \cite{Gogolin:2016} for a review).
Such equilibration can be rigorously shown to happen if the spectrum of the Hamiltonian fulfills certain non-resonance conditions, and the initial state has overlap with many energy eigenstates or the second most populated eigenstates is occupied with only a small probability (a weaker requirement).
In these results, the equilibrium expectation values or reduced states are given by the diagonal ensemble (also known as infinite time averaged state, dephased state, or pinched state).
How and to which state equilibration occurs, of course is closely connected to whether a system is ergodic or not, which motivates us to consider in more detail the correlations in this diagonal ensemble to study ergodicity breaking.

Concepts of quantum information have been useful in the typicality approach (system-bath entanglement) \cite{Popescu:2006}, the dynamical equilibration approach, and in the ETH approach, for example in studying the volume law scaling of entanglement in eigenstates, the crossover to an area law is a signature of the MBL transition \cite{Luitz:2015}. In a recent work, some of the current authors have proposed a different information theory inspired approach~\cite{Goold:2015}.
The idea is to look at the correlations within the diagonal ensemble to understand non-ergodic behavior in the context of the MBL transition \cite{Goold:2015}. The purpose of the current work is to demonstrate that this concept is more generally useful and can detect ergodicity breaking in a range of scenarios beyond and including MBL.
The formalism offers a fresh approach to ergodicity and its breaking in quantum systems, while at the same time giving us novel insights into the structure of correlations in the equilibrium state of dynamical systems.

\section{Ergodicity and total correlations}
Due to the absence of a universally valid phase space picture in quantum systems it is not obvious how to generalize the concept of ergodicity to the quantum realm, especially in systems that do not have a well defined classical limit.
As was outlined in Ref.~\cite{Goold:2015}, the total correlations in the diagonal ensemble offer a physically meaningful way to define and probe ergodicity and its breaking in quantum systems.
Here we generalize this approach.

\subsection{A condition for ergodicity.}
The (quasi) ergodic hypothesis in classical systems states that over time a system's dynamics uniformly covers its entire phase space so that the (infinite time) time average and the micro-canonical averages agree~\cite{Gallavotti:2013}.
It is thus natural to define ergodicity in quantum systems in an analogous way via the portion of the explored Hilbert space.
A complication is that quantum systems explore all of the available phase space uniformly such that time and microcanonical average agree exactly only for very special initial states.
This naturally leads us to build a notion of ergodicity in quantum systems based on the fraction of the available Hilbert space that is explored, as opposed to classical notion of ergodicity that requires that all of the available phase space is explored uniformly.
The available Hilbert space hereby can be usually naturally defined as, for example, the fixed magnetization or fixed filling fraction subspace if the system has such symmetries.
To define ergodicity via the fraction of Hilbert space that is explored one obviously first needs to devise a way of quantifying the explored fraction.
It is this question that we elucidate in this work, going beyond the initial proposal in Ref.~\cite{Goold:2015}.

For a fixed initial state $\rho$ and non-degenerate Hamiltonian $H$ the diagonal ensemble is defined as 
\begin{equation} \label{eq:timeav}
  \omega \coloneqq  \sum_n \ket {E_{n}}\bra {E_{n}} \,\rho\, \ket {E_{n}}\bra {E_{n}} = \lim_{\tau\to\infty} \frac{1}{\tau} \int^{\tau}_{0}dt\, \e^{-itH}\, \rho\, \e^{itH} ,
\end{equation}
where $\ket {E_{n}}$ are the eigenvectors of $H$.
The state $\omega$ is the state that maximizes the von Neumann entropy subject to all constants of motion \cite{PhysRevLett.10-6}.
For pure initial states $\rho=|\Psi\rangle\langle\Psi|$, the inverse purity $1/\tr(\omega^2)$ of the diagonal ensemble can be seen as a measure for how spread-out the initial state was over the different eigenstates of the Hamiltonian and it often goes under the apt name of \emph{effective dimension} or \emph{participation ratio}.
If the effective dimension is high, expectation values of observables can be rigorously shown \cite{Reimann:2008,Linden:2009,Polkovnikov:2011,Eisert:2015,Gogolin:2016} to equilibrate on average during the time evolution towards their values in the state $\omega$.
The effective dimension however is not the only way of quantifying the spreading of the initial state.
Another measure is the von Neumann entropy of the diagonal ensemble $S(\omega)=-\sum_n p_n\log_2 p_n$ with $p_n=|\langle E_n|\Psi\rangle|^2$ and derived quantities; this was the route taken in Ref.~\cite{Goold:2015}.

Both of these quantities are special cases of a whole family of entropies, the so-called $\alpha$-Renyi entropies, which for $0<\alpha<\infty$ are defined as
\begin{equation}\label{eq:renyidiagentrop}
  S_\alpha(\rho) \coloneqq \frac{1}{1-\alpha} \log_2(\tr(\rho^\alpha)) .
\end{equation}
The $2$-entropy $S_2 = \log_2(1/\tr(\rho^2))$ is the log of the inverse purity, the $1$-entropy $S_1 = S$ is the von~Neumann entropy, and the two extreme cases are defined as $S_\infty(\rho) \coloneqq \log_2(1/\|\rho\|_\infty)$ and $S_0(\rho) \coloneqq \log_2(\rank(\rho))$.
The Renyi entropies are monotonically non-increasing as a function of $\alpha$.
In other words, for any fixed $\rho$, it holds that $S_\alpha(\rho) \leq S_{\alpha'}(\rho)$ whenever $\alpha' \geq \alpha$.
For $\psi$ a pure state and $\rho$ a normalized quantum state of a system with Hilbert space dimension $d$, it holds that
\begin{equation}
  0 = S_\alpha(\psi) \leq S_\alpha(\rho) \leq S_\alpha(\1_{d\times d}/d) = \log_2(d) .
\end{equation}
Except in the case $\alpha = 0$ the inequalities hold with equality only if $\rho$ is either pure or maximally mixed, respectively.

It is the upper bound that interests us here.
Given a Hamiltonian $H$, an initial state $\rho$ explores all of Hilbert space if $S_\alpha(\omega) = \log_2(d)$.
As said above, this however only happens for very special states for which $\omega$ is maximally mixed.
A natural relaxation of this condition is to demand that for some chosen $0\leq\alpha\leq\infty$ there exists a constant $\lambda > 0$, independent of $N$ and $d$, such that $S_\alpha(\omega) \geq \log_2(\lambda\,d)$, i.e., that the state explores a $\lambda$-fraction of the Hilbert space as measured by the $\alpha$-Renyi entropy.

That this is a sensible condition for ergodicity is further illustrated by the following consideration:
For any fixed initial state $\rho$ and Hamiltonians $H$ with eigenbasis randomly drawn from a unitary invariant ensemble on a Hilbert space of dimension $d$ one can show for the $\alpha=1$ entropy \cite[Eq.~(B6)]{Hayden:2004} and the $\alpha=2$ entropy \cite{Linden:2009} that the probability that the state explores less than half of the available Hilbert space is at least almost exponentially suppressed with growing $d$ (as the Renyi entropies are non-increasing as a function of $\alpha$ this then holds for all $0 \leq \alpha \leq 2$).
That is, any fixed initial state is with high probability ergodic according to our condition with respect to Hamiltonians drawn unitarily at random---as one would expect. To be more precise: Generalizing the considerations from \cite{Goold:2015} we hence demand that a system should be considered $\alpha$-ergodic only if the initial states explore at least a constant fraction of the available Hilbert space in the sense that for some $\lambda$ it holds that $S_\alpha(\omega) \geq \log_2(\lambda\,d)$.  In the models that we will consider the Neel states are suitable initial states.

We have so far defined a family of conditions parametrized by $\alpha$ that appear as natural quantum generalizations of the concept of ergodicity, but have not yet said much about the role of $\alpha$. 
Remember that the Renyi entropies are monotonically non-increasing as a function of $\alpha$.
Thus, demanding that, for example, $S_2 \geq \log_2(\lambda\,d)$ is a stronger requirement than demanding that the same scaling holds for $S_1$.
As we will see later, the fact that a system fulfills our condition for ergodicitiy for a given $\alpha$ has direct consequences on the scaling of the total correlations with the number of particles.

\subsection{Total Correlations.}
Phase transitions that involve the breaking of ergodicity, like the MBL transition, have in the past been analyzed with various measures of correlations.
A focus thereby was on the mutual information, which was found to saturate to a constant in Anderson localization, grow logarithmically in time in the MBL phase, and linearly in ergodic phases \cite{DeTomasi2016}.
Further, it decays exponentially with the distance between subsystems in the localized phase, but slower than exponential in the ergodic phase. Here we concentrate on a correlation measure called the total correlations and its Renyi generalizations.
Concretely we define the $\alpha$-Renyi total correlations as
\begin{equation}
  \T_\alpha(\rho) \coloneqq \sum_{m=1}^N S_\alpha(\rho_{m}) - S_\alpha(\rho),
  \label{eq:totcor1}
\end{equation}
where $\rho_m$ is the marginal (reduced state) of $\rho$ on site $m$.
In the special case $\alpha=1$ the total correlations have the following operational meaning:
Let $\mathcal{P}$ be the set of all product states of an $N$ partite quantum system, i.e., for spin systems, states of the form $\pi=\pi_{1}\otimes\pi_{2}\dots\otimes\pi_{N}$ and the obvious analogues for fermionic and bosonic systems, then \cite{Modi:2010} 
\begin{equation}
  \T_1(\rho) = \min_{\pi\in\mathcal{P}} S(\rho\|\pi) .
\end{equation}
where $S(\rho\|\sigma) \coloneqq -\tr(\rho\,\log_2 \sigma)-S_1(\rho)$ is the relative entropy between the states $\rho$ and $\sigma$ and it can be thought of as a measure of distinguishability of the two states.
More precisely, the relative entropy quantifies how difficult it is to distinguish between many copies of $\rho$ and many copies of $\sigma$ in a hypothesis testing scenario \cite{Audenaert2012}.
It turns out that there is a unique product state that minimizes the relative entropy in the above expression and this is the product of the reduced states $\rho_{m}$ of $\rho$, i.e., $\pi=\otimes^{N}_{m=1} \rho_{m}$ \cite{Modi:2010}.
In the case $\alpha =1$ the total correlations, can hence be thought of as the distinguishability from the closest product state.
No such straightforward operational interpretation exists for $\alpha\neq1$ to the best of our knowledge. 

As we will explain in the next sections, insights into integrability breaking can be obtained through the various Renyi total correlations.

\subsection{Scaling of the total correlations.}
In the following we analyze the total correlations, and in particular $T_1$ and $T_2$, of the diagonal ensemble $\omega$ for Neel initial states in various spin chain models.
One characteristic that will be very insightful is the scaling of the total correlations with $N$.

Inspecting Eq.~\eqref{eq:totcor1} one might expect that the total correlations $\T(\omega)$ in the diagonal ensemble should generally scale extensively in the system size $N$, i.e, for large $N$, to leading order, it should scale like
\begin{equation} \label{eq:linearscaling}
  \T_\alpha(\omega) \propto N ,
\end{equation}
as $\T_\alpha(\omega)$ involves the sum $\sum_{m=1}^N S_\alpha(\omega_{m})$ of the $N$ subsystem entropies.

If a family of systems of increasing size satisfies the condition for ergodicity defined above, then the contribution linear in $N$ from the first sum can be precisely canceled by the $-S_\alpha(\omega)$ term; $S_\alpha(\omega)$ is known as the diagonal entropy~\cite{PolkovnikovReview} and is a measure of localization in the energy eigenbasis.
Consider a quantum spin chain of local dimension $2$ in the zero magnetization subspace~\footnote{The diagonal entropy - or Shannon entropy in the energy eigenbasis, measures delocalization/localization in that basis and therefore its scaling can give information about ergodicity. However, the total correlations and its Renyi generalizations give far more information; in particular it not only scales differently in different phases but also shows divergence of the correlations in the critical region - an interesting phenomenon in its own right which cannot be studied just looking at the diagonal entropy alone.}.
The available Hilbert space dimension is $d = \binom{N}{N/2} = N! / \left(\tfrac{N}{2}!\right)^{2} \geq \sqrt{8\,\pi}\, \e^{-2}\, 2^N/\sqrt{N}$ and $S_\alpha(\omega_{m}) \leq \log_2 2 = 1$, so that if the condition for ergodicity $S_\alpha(\omega) \geq \log_2(\lambda\,d)$ holds, one finds at most the logarithmic scaling
\begin{equation} \label{eq:logscaling}
  \T_\alpha(\omega) \leq \log_2(N)/2 - \log_2(\lambda\,\sqrt{8\,\pi}\,\e^{-2}) .
\end{equation}
One furthermore retains a logarithmic scaling for ergodic systems for all other constant magnetization/fillings subspaces $\eta\neq 1/2$ in the case $\alpha=1$ \cite{Goold:2015}.

This sub-extensive scaling can also be understood intuitively: The transport present in ergodic systems correlates the different parts of the system to the extent that they appear, for most times during the evolution, so mixed that the time averaged state starts to resemble a product state.

In conclusion we can say that whenever we see a faster than logarithmic scaling in the $\alpha$-Renyi total correlations of the diagonal ensembles $\omega$ of an initial state from the half filling subspace, then the condition for ergodicity for that value of $\alpha$ is violated.
On the contrary, a logarithmic scaling suggests ergodic behavior.

\section{Examples}
Low-dimensional many-body quantum systems, such as spin-1/2 chains are systems commonly used to study ergodicity breaking phenomena.
In what follows we will always consider dynamics starting from the trivial initial Hamiltonian, $H_{0}$ defined as \begin{equation} \label{eq:trivial}
  H_{0} =\sum^{N}_{i=1}J_zs^{i}_{z}s^{i+1}_{z},\end{equation}
where $s^i$ are spin-1/2 operators, and we choose as an initial state the Neel state $|\Psi_{0}\rangle=|\uparrow\downarrow\uparrow\downarrow\uparrow\downarrow\dots\rangle$.
Now imagine a quench where we turn on additional terms denoted by an interaction part $H_{int}$ such that dynamics is initiated and governed by the Hamiltonian $H_{F}=H_{0}+H_{int}$.
We shall build the diagonal ensemble defined by Eq.~\eqref{eq:timeav} by exact diagonalization and then investigate the scaling of the total correlations $T(\omega)$ as defined in Eq.~\eqref{eq:totcor1}, both in the case of the Von Neumann total correlations ($\alpha=1$) and of the $2$-Renyi total correlations ($\alpha=2$).
We choose the initial state $|\Psi_{0}\rangle$ as a Neel state for two principal reasons:
First, in the models we shall consider, the Neel state can be shown to sample eigenstates of $H_F$ at the center of the spectrum and in the half filling subspace~\cite{torresherrera2014}.
In this regime we expect the finite size effects to be minimized.
Second, the Neel state (charge density wave in fermion picture), is by now routinely prepared by experimentalists to study ergodicity breaking, for example in the recent studies of MBL systems~\cite{Schreiber:2015,Smith:2016}.

\subsection{Integrability Breaking}
\begin{figure}[t]
\includegraphics[width=1\columnwidth]{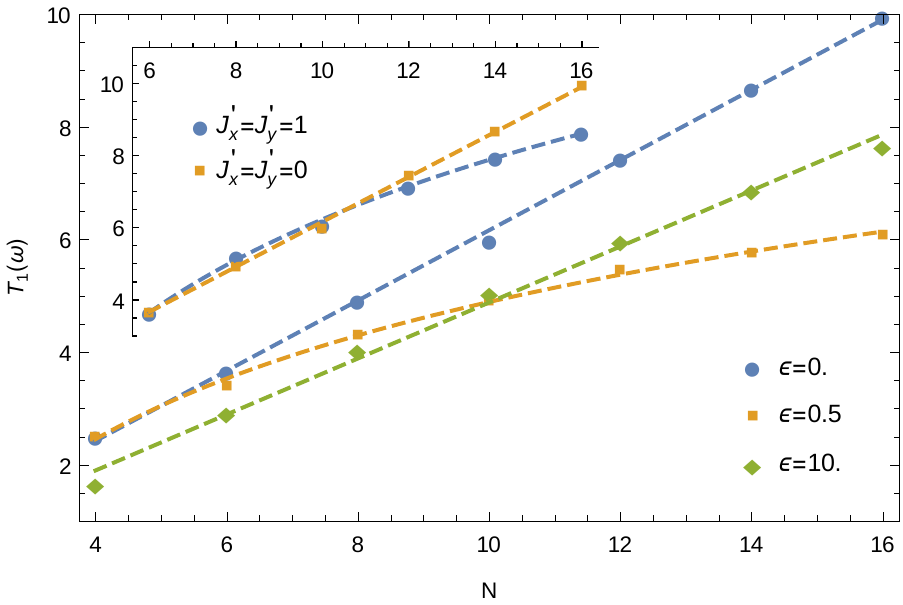}
\caption{(Color online) The Von Neumann total correlations of the diagonal ensemble starting with the Neel state for an XXZ chain with defect of strength $\epsilon$ placed at centre of the chain (Eq.
\eqref{eq:heisenbergchain} with parameters $J_x=J_y=1$ and $J_z=0.5$).
When the defect strength is zero or very strong the model is integrable, which is reflected in a linear scaling of the total correlations, and when it is comparable with the interaction energy it shows a logarithmic growth indicative of ergodic dynamics.
\textit{Inset. } Total correlations for an XXZ chain with next-nearest-neighbour interaction (Eq.
\eqref{eq:nnn} with parameters $J_x=J_y=1$, $J_z=0.5$ and $J_x'=J_y'=1$, compared to the same model with $J_x'=J_y'=0$).
The model is non-integrable and thus the scaling of the total correlations is logarithmic in the system size.}
  \label{fig1}
\end{figure}

\begin{figure}
\includegraphics[width=1\columnwidth]{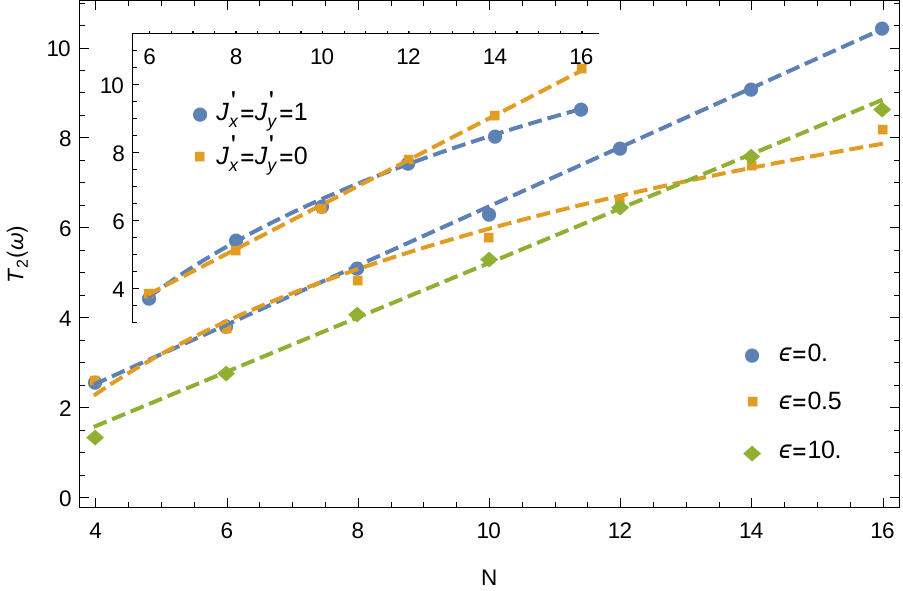}
\caption{(Color online) The $2$-Renyi total correlations of the diagonal ensemble starting with the Neel state for an XXZ chain with defect of strength $\epsilon$ placed at centre of the chain (Eq.
\eqref{eq:heisenbergchain} with parameters $J_x=J_y=1$ and $J_z=0.5$).
When the defect strength is zero or very strong $T_2$ scales linearly with system size and when it is comparable with the interaction energy it scales logarithmically.
\textit{Inset. } Total correlations for an XXZ chain with next-nearest-neighbour interaction (Eq.
\eqref{eq:nnn} with parameters $J_x=J_y=1$, $J_z=0.5$ and $J_x'=J_y'=1$, compared to the same model with $J_x'=J_y'=0$).
The scaling of the total correlations is logarithmic in the system size.}
  \label{fig1sp}
\end{figure}

Let us begin with the following model studied by Santos in 2004 \cite{Santos:2004}.
The model is an XXZ spin chain with open boundary conditions which includes a single defect at the centre of the chain of strength $\epsilon$,
\begin{equation} \label{eq:heisenbergchain}
  H_{F} =\sum^{N}_{i=1}\Big[J_xs^{i}_{x}s^{i+1}_{x}+J_{y}s^{i}_{y}s^{i+1}_{y}+J_zs^{i}_{z}s^{i+1}_{z}\Big]+\epsilon s^{N/2}_{z}.
\end{equation}
The integrability of the chain is broken\cite{Santos:2004}, indicated by a crossover from Poissonian to Wigner-Dyson statistics, for defect strengths which are comparable to the interaction energy.
As the strength of the single defect is increased the system becomes integrable again as the chain is cut into two XXZ chains.
Our theory predicts then that we should see a linear-log-linear behavior in the scaling of the total correlations as we increase the defect strength from zero.
This is indeed what results from the numerical computation of $T_1(\omega)$, shown in the main plot of Fig.~\ref{fig1} for three values of $\epsilon$, $\epsilon=0$, $0.5$ and $10$: $T_1$ scales linearly for the values $\epsilon=0$ and $\epsilon=10$ and approximately logarithmically for $\epsilon=0.5$ as a function of system size. The same happens to the $2$-Renyi total correlations $T_2(\omega)$, shown in the main plot of Fig.~\ref{fig1sp}.

The second model that we consider is the clean XXZ model with next-nearest-neighbour interaction,
\begin{multline} \label{eq:nnn}
  H_{\text{NNN}} =\sum^{N}_{i=1}\Big[J_x s^{i}_{x} s^{i+1}_{x}+J_{y} s^{i}_{y} s^{i+1}_{y}+J_z s^{i}_{z} s^{i+1}_{z}+\\ J_x' s^{i}_{x} s^{i+2}_{x}+ J_{y}' s^{i}_{y} s^{i+2}_{y}\Big].
\end{multline}
For $J_x', J_y'\neq0$ integrability is broken and the scaling of the total correlations $T_1(\omega)$ with the system size is logarithmic, as is shown in the inset of Fig.~\ref{fig1}.
Also in this model we see an analogous behaviour for the $2$-Renyi total correlations $T_2(\omega)$, which are logarithmically scaling with the system size (see the inset of Fig.~\ref{fig1sp}).

In both cases of integrability breaking, the total correlations displays the predicted behaviour.

\subsection{Many-body localization} 
\begin{figure}[t]
\includegraphics[width=1\columnwidth]{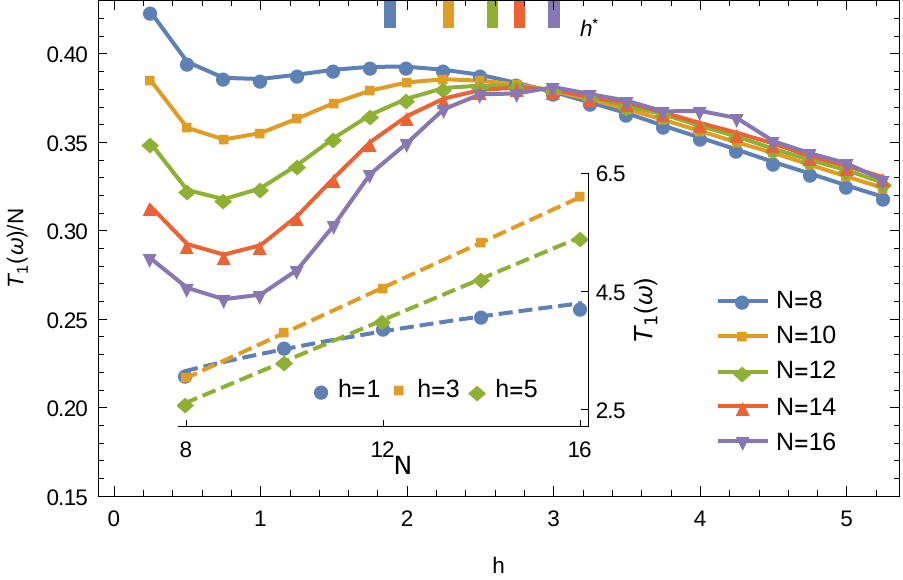}
\caption{(Color online) The Von Neumann total correlations of the diagonal ensemble, starting with a Neel state, for the Heisenberg model with random fields (whose Hamiltonian is Eq.
\eqref{eq:mbl} with $h_i\in[-h,h]$ and $J_x=J_y=J_z=1$), rescaled with the system size. The markers on the top axis denote the positions of the local peak $h^*(N)$.
The curves show a system-size-dependent peak (see Fig.~\ref{fig:peaksp}) and collapse for $h\gtrsim2.5$.
\textit{Inset.
} System size scaling of the total correlations for three example values of $h$, showing a logarithmic scaling deep in the delocalized phase and a linear scaling for disorder values near the transition (which is at $h_c\approx3.7$) and in the localized phase.}
  \label{fig2}
\end{figure}

\begin{figure}[t]
\includegraphics[width=1\columnwidth]{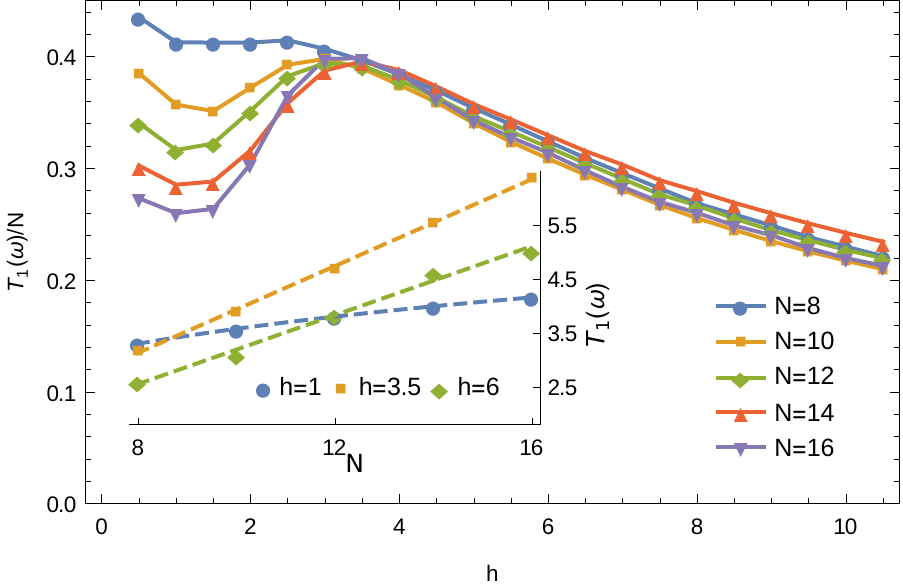}
\caption{(Color online) The Von Neumann total correlations of the diagonal ensemble, starting with a Neel state, for the Aubry-Andr\'e model (whose Hamiltonian is Eq.
\eqref{eq:mbl} with the cosine $h_i$ fields and $J_x=J_y=J_z=1$), rescaled with the system size.
The curves show a system-size-dependent peak and collapse for $h\gtrsim3.5$.
\textit{Inset.
} System size scaling of the total correlations for three example values of $h$, showing a logarithmic scaling deep in the delocalized phase and a linear scaling in the localized phase.}
  \label{fig3}
\end{figure}

Let us now consider models which have an MBL transition that separates an ergodic phase and a non-ergodic one where a sufficient number of local integrals of motion exists in order to have a breaking of the ETH~\cite{ros2015integrals}.
We look at a system with the Hamiltonian
\begin{equation} \label{eq:mbl}
  H_{\text{MBL}} =\sum^{N}_{i=1}\Big[J_x s^{i}_{x} s^{i+1}_{x}+J_{y} s^{i}_{y} s^{i+1}_{y}+J_z s^{i}_{z} s^{i+1}_{z}+h_i s^i_z\Big],
\end{equation}
where $h_i\in[-h,h]$ is a disordered field (Heisenberg model with random fields) or $h_i=h\cos{(2\pi \phi^{-1} i+\delta)}$, where $\phi$ is the golden ratio and $\delta$ is a random phase in $[0,2\pi)$, that is a pseudo-disordered cosine field (Aubry-Andr\'e model).
For both models we compute the total correlations for the diagonal ensemble with the Neel initial state, averaging over many disorder or pseudo-disorder realizations, the latter obtained through the random phase $\delta$ ($10^5$ realizations for $N\leq 12$, $10^4$ for $N=14$ and $250$-$1000$ for $N=16$).
The results for the Von Neumann total correlation rescaled with the system size, $T_1(\omega)/N$, are shown in Figs.~\ref{fig2} and \ref{fig3} respectively for the two models, as a function of the disorder or quasi-disorder strength.

We note two features: the curves collapse for $h\geq h^*$, indicating a linear scaling of the total correlations and thus non-ergodicity, and $T_1(\omega)/N$ peaks at a value $h^*(N)$.
Remarkably, the presence of a peak can be understood as a divergence of correlations at the MBL transition point and its asymptotic position in the infinite-size limit gives the transition value $h_c$~\cite{Goold:2015}.
We are able to perform such extrapolation (see Fig.~\ref{fig:peaksp}), thus obtaining $h^\text{H}_c= \lim_{N\to\infty} {h^\text{H}}^*=4.0\pm0.2$ for the random potential and $h^\text{AA}_c= \lim_{N\to\infty} {h^\text{AA}}^* =4.5\pm0.9$ for the Aubry-Andr\'e potential. Note that for the latter case the extrapolation suffers from much larger errors due to the smaller movement of the peak of the finite-size data with respect to its error.

For the Heisenberg model with random fields the transition value has been estimated through other numerical evidence \cite{Luitz:2015,Pal:2010,pietracaprina2016forward} to be equal to $h^\text{H}_c=3.7(2)$ at the center of the band for the parameters that we used, although its actual value could be larger ($h^\text{H}_c\geq 4.5$ according to~\cite{Devakul2015}); an equivalent high-quality numerical result is not available for the Aubry-Andr\'e model, although experimental works find the localization transition at similar values~\cite{Schreiber:2015}.
Interestingly, as soon as interactions are introduced, the Aubry-Andr\'e model acquires almost identical features to the Heisenberg with random fields model first studied by total correlations in \cite{Goold:2015}.
Finally, for both models, for weak \mbox{(quasi-)} disorder ($h\lesssim h^*$), the scaling of the total correlations is logarithmic, implying an ergodic phase.
 
\begin{figure}
\includegraphics[width=1\columnwidth]{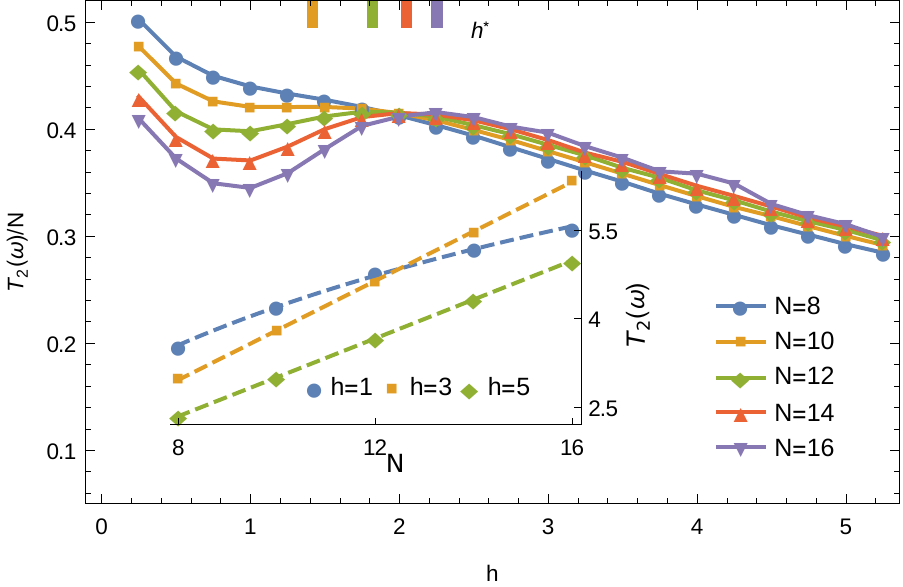}
\caption{(Color online) The $2$-Renyi total correlations of the diagonal ensemble, starting with a Neel state, for the Heisenberg model with random fields (whose Hamiltonian is Eq.
\eqref{eq:mbl} with $h_i\in[-h,h]$ and $J_x=J_y=J_z=1$), rescaled with the system size. The markers on the top axis denote the positions of the local peak $h^*(N)$, excluding the case $N=8$, where no local maximum can be discerned.
The curves show a system-size-dependent peak (see Fig.~\ref{fig:peaksp}) and collapse for $h\gtrsim2$.
\textit{Inset.
} System size scaling of the total correlations for three example values of $h$, showing a logarithmic scaling deep in the delocalized phase and a linear scaling for disorder values near the transition (which is at $h_c\approx3.7$) and in the localized phase.}
  \label{fig2sp}
\end{figure}

Let us now consider the $2$-Renyi total correlations, focusing first on the Heisenberg model with random fields. The total correlations rescaled by the system size $T_2(\omega)/N$ are shown in Fig.~\ref{fig2sp}, showing a collapse for $h\gtrsim 2$ for the available system sizes, to be compared with an analogous behaviour of $T_1(\omega)/N$, where the collapse point is $\approx 2.5$.
For a system of infinite size, one would expect that the collapse points (or equivalently the peak positions) should be the same for all total correlations $T_\alpha$; due to the stronger ergodicity requirement of higher-$\alpha$ Renyi entropies, however, it is understandable that $T_2$ gives an underestimation at finite, small system sizes.

As expected, $T_2$ scales linearly with the system size for $h\gtrsim h^*_2$ and logarithmically for $h\lesssim h_2^*$.
Moreover, the curves in Fig.~\ref{fig2sp} peak on a system-size-dependent value $h_2^*(N)$.
In Fig.~\ref{fig:peaksp} we show the scaling of the $T_1$ and $T_2$ peak positions with the system size.
For both the $T_1$  and $T_2$ peak positions, the finite-size scaling is very well approximated by a linear behaviour in $1/N$; the infinte-size extrapolation for the $2$-Renyi case is $h^*_2(\infty)=3.6\pm0.2$, which is lower than the value obtained from the Von Neumann total correlations. This is again a signal of the underestimation of the breaking of ergodicity and of stronger finite size effects due to the hierarchy in the Renyi entropies.

\begin{figure}
\includegraphics[width=1\columnwidth]{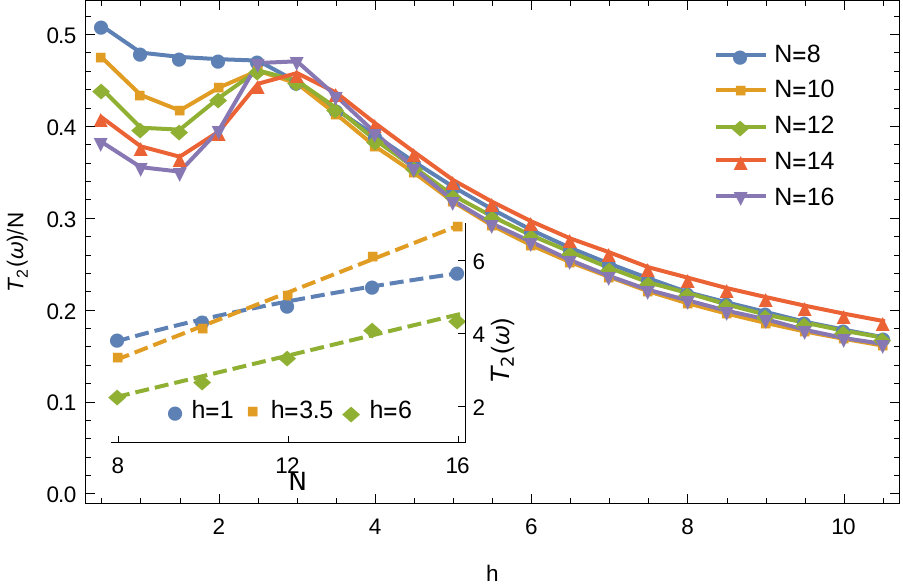}
\caption{(Color online) The $2$-Renyi total correlations of the diagonal ensemble, starting with a Neel state, for the Aubry-Andr\'e model (whose Hamiltonian is Eq.
\eqref{eq:mbl} with the cosine $h_i$ fields and $J_x=J_y=J_z=1$), rescaled with the system size.
The curves show a system-size-dependent peak (see Fig.~\ref{fig:peaksp}) and collapse for $h\gtrsim 3$.
\textit{Inset.
} System size scaling of the total correlations for three example values of $h$, showing a logarithmic scaling deep in the delocalized phase and a linear scaling in the localized phase.}
  \label{fig:tc2aam}
\end{figure}

Finally, in Fig.~\ref{fig:tc2aam} we show the results for the $2$-Renyi total correlations of the Aubry-Andr\'e model. The qualitative behaviour is again the same as for the model with random fields, showing once again that the interactions remove the special integrability features of a non-interacting Aubry-Andr\'e model.  Specifically, $T_2(\omega)/N$ has a peak which scales to $h_2^*(\infty)=3.5\pm0.6$, which, analogously to what happens in the random fields model, is a lower value than the result for the extrapolated peak of the rescaled Von Neumann total correlations.

\begin{figure}
\includegraphics[width=1\columnwidth]{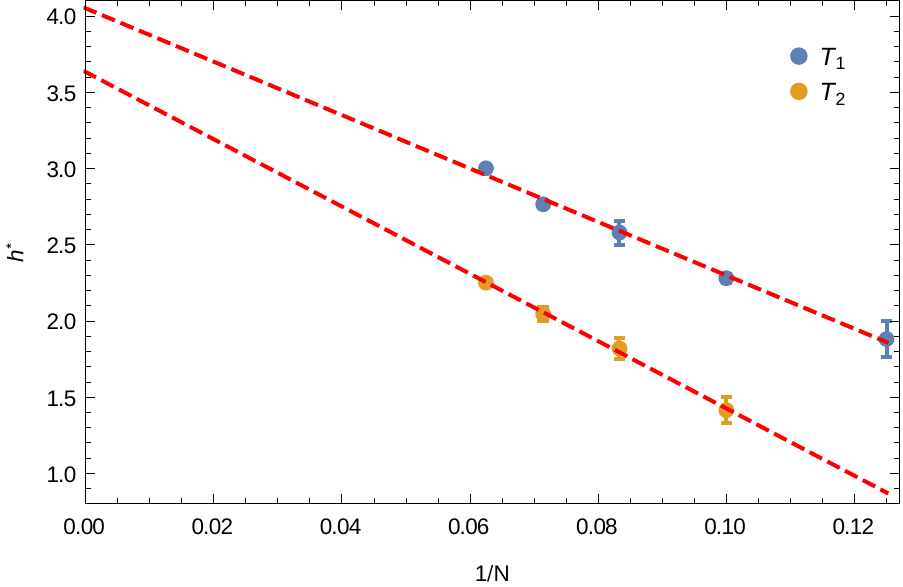}
\caption{(Color online) System size scaling of the peak of the total correlations in the Heisenberg model with random fields, for both $T_1$ and $T_2$.
The peak is extracted from a polynomial interpolation of each of the curves in Figs.~\ref{fig2} and \ref{fig2sp}, in which it is denoted with a marker on the top axis.
We perform a linear fit in $1/N$ and obtain the infinite size extrapolation given in the text.}
  \label{fig:peaksp}
\end{figure}

\section{Conclusions}
The results displayed in this work demonstrate that the total correlations of the diagonal ensemble is a powerful concept to understand ergodicity breaking in quantum systems in general.
The numerics performed in different models confirm that the scalings predicted in the theory first outlined in \cite{Goold:2015} work in the cases of both integrability breaking and also ergodic to MBL transitions.
Within the context of systems which show ergodicity breaking, a number of methods have proven useful, especially in the case of localizable systems, consisting in examining the participation ratio~\cite{de2013ergodicity}, the entanglement entropy~\cite{Bardason2012} and the full entanglement spectrum~\cite{entspect2016}. The total correlations presented here are an additional tool for such systems.
Where MBL systems are concerned the rescaled quantity offers the additional feature of peaking around the expected transition point; given that this may be seen as a divergence of correlations in the infinite time steady state, it represents a novel contribution to the theory of quantum correlations. The peak was observed in two models displaying a MBL transition.

We have examined the total correlations \eqref{eq:totcor1} for two values of $\alpha$, $\alpha=1$ and $\alpha=2$; the $2$-Renyi total correlations has the numerical advantage, being more suitable to be computed avoiding the diagonalization of the density matrix through faster techniques such as t-DMRG, where a finite time average could be performed.
As future work we intend to increase the system sizes dramatically by applying t-DMRG and finite time averaging to compute the $2$-Renyi total correlations.

Finally, given the rapid progress in experimental techniques it seems possible that the total correlations could be experimentally measured in the near future.
The total correlations amounts to subtracting the diagonal entropy from the sum of the entropies of the marginal states. 
The marginal states and their entropies can already be measured in a quench starting from a Neel state \cite{Smith:2016}.
Measuring the diagonal entropy is a more challenging task, however, some progress has been made in this direction for small systems \cite{Neill:2016} and given the current interest in measuring Renyi entropies in experiments \cite{Islam:2015} we believe that further advances could yield the experimental extraction of total correlations.

\section*{Acknowledgements}
FP would like to thank G. Parisi for useful discussions and SISSA and ICTP for access to computing resources and hospitality during the completion of this work.
CG acknowledges support from MPQ-ICFO, ICFOnest+ (FP7-PEOPLE-2013-COFUND), and co-funding by the European Union's Marie Sk\l{}odowska-Curie Individual Fellowships (IF-EF) programme under GA: 700140, as well as from the European Research Council
(ERC AdG OSYRIS and CoG QITBOX), Axa Chair in Quantum Information Science, 
The John Templeton Foundation, Spanish MINECO (FOQUS 
FIS2013-46768 and Severo Ochoa Grant No.~SEV-2015-0522),
Fundaci\'{o} Privada Cellex, and Generalitat de Catalunya (Grant No.~SGR 874 and 875).
\bibliography{ergodic}

\end{document}